\documentclass{Interspeech2024}
\usepackage{multirow}
\usepackage{tabularx}
\usepackage{booktabs}
\usepackage{hyperref}
\usepackage{url}
\usepackage{xcolor}
\PassOptionsToPackage{hyphens}{url}
\usepackage{hyperref}

\newcommand{\kedit}[1]{\textcolor{black}{#1}} 
\newcommand{\kkedit}[1]{\textcolor{black}{#1}} 
\newcommand{\kkkedit}[1]{\textcolor{black}{#1}} 
\newcommand{\ryedit}[1]{\textcolor{black}{#1}} 
\newcommand{\rrryedit}[1]{\textcolor{black}{#1}} 
\newcommand{\hedit}[1]{\textcolor{black}{#1}} 
\newcommand{\sedit}[1]{\textcolor{black}{#1}} 

\interspeechcameraready

\title{LibriTTS-P: A Corpus with Speaking Style and Speaker Identity Prompts for Text-to-Speech and Style Captioning}

\name[affiliation={1}]{Masaya}{Kawamura}
\name[affiliation={1}]{Ryuichi}{Yamamoto}
\name[affiliation={1}]{Yuma}{Shirahata}
\name[affiliation={1}]{Takuya}{Hasumi}
\name[affiliation={1}]{Kentaro}{Tachibana}

\address{
  $^1$LY Corp., Japan
}
\email{kawamura.masaya@lycorp.co.jp}


\begin{document}

\maketitle
\begin{abstract}
We introduce LibriTTS-P, a new corpus based on LibriTTS-R that includes utterance-level descriptions (i.e., \textbf{p}rompt\sedit{s}) of speaking style and speaker-level prompts of speaker characteristics. We employ a hybrid approach to construct prompt annotations: (1) manual annotations that capture human perceptions of speaker characteristics and (2) synthetic annotations on speaking style. Compared to existing English prompt datasets, our corpus provides more diverse prompt annotations for all speakers of LibriTTS-R.
Experimental results for prompt-based controllable TTS demonstrate that the TTS model trained with LibriTTS-P achieves higher naturalness than the model using the conventional dataset.
Furthermore, the results for style captioning tasks show that the model utilizing LibriTTS-P generates 2.5 times more accurate words than the model using a conventional dataset. 
\kkkedit{Our corpus, LibriTTS-P, is available at \url{https://github.com/line/LibriTTS-P}.}

\end{abstract}
\vspace{-2mm}
\section{Introduction}
\label{sec:intro}


Recently, prompt-based controllable \ryedit{text-to-speech (TTS)} models have attracted considerable interest~\cite{guo2023prompttts,liu2023promptstyle,vyas2023audiobox,lyth2024natural,zhang2023promptspeaker,sigurgeirsson2024controllable}.
These models use a natural language description (\ryedit{referred to as a prompt}) to control voice characteristics.
Furthermore, recent years have seen research into the models that generate natural language descriptions of speaking style from speech~\cite{yamauchi2024stylecap}.

Towards research on prompt-based speech applications, previous studies have constructed datasets with annotated prompts for TTS datasets~\cite{guo2023prompttts,yang2023instructtts,ji2023textrolspeech}.
For instance, Guo et al.~\cite{guo2023prompttts} present PromptSpeech, which comprises annotated text prompts about five style factors: gender, speaking speed, pitch, loudness, and emotion, for a TTS-generated dataset and a small subset of LibriTTS~\cite{zen2019libritts}.
However, since most previous works focus on prompts about speaking style, they are limited in describing the rich characteristics of human speech, such as speaker identity.
To overcome this problem, PromptTTS++~\cite{shimizu2024prompttts} utilizes an additional prompt describing human perceptions of speaker identity. 

On the other hand, there has been interest in utilizing web-crawled data for prompt annotations~\cite{watanabe2023coco-nut}. 
The large-scale crawled data can cover much richer voice characteristics than the \kkedit{existing prompt-based TTS} datasets.
However, it is often challenging to obtain studio-quality recordings that are necessary for achieving high-quality prompt-based TTS.

\begin{figure}[!t]
  \centering
  \includegraphics[width=\linewidth]{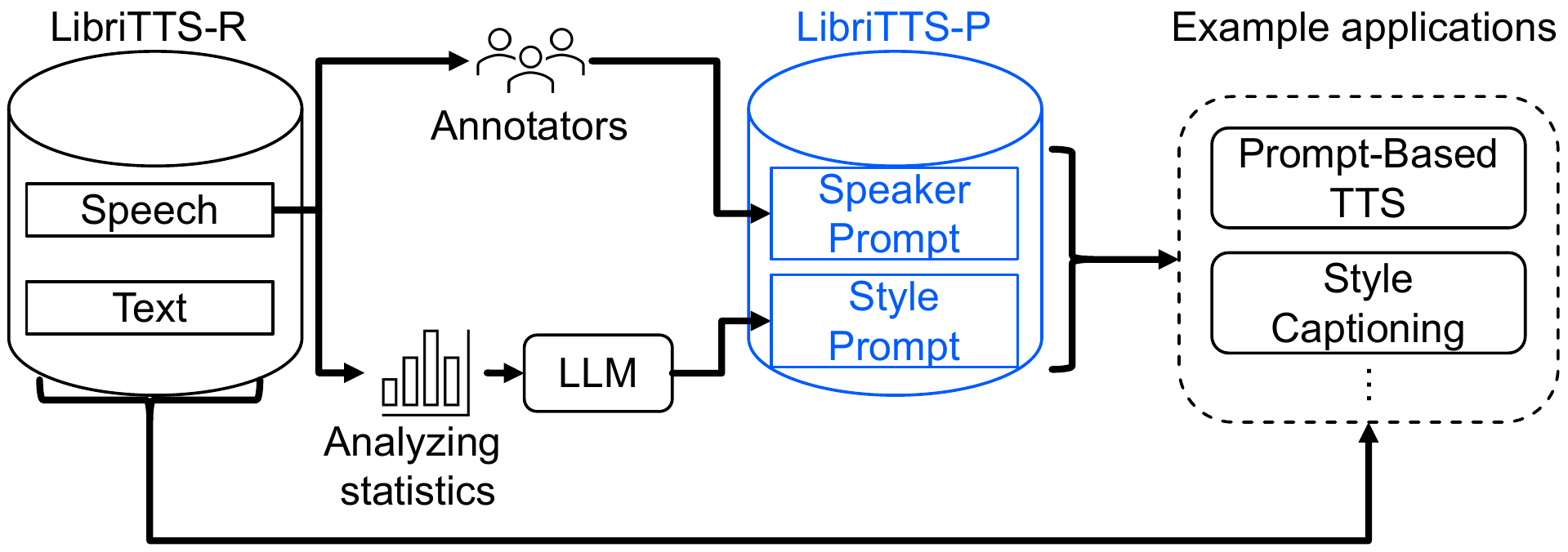}
  \vspace{-4mm}  
  \caption{Overview of LibriTTS-P \ryedit{and its applications}.}
  \vspace{-6mm}
  \label{fig:libritts_p_overview}
\end{figure}


This work aims to provide prompt annotations for an existing high-quality TTS dataset with rich text prompt variations.
To this end, we present LibriTTS-P, a new corpus based on LibriTTS-R that contains text prompts describing speaking style and speaker identity. 
Figure~\ref{fig:libritts_p_overview} shows the overview of LibriTTS-P \ryedit{and its applications}.
Additionally, Table~\ref{tab:dataset_comparison} compares our corpus with the most relevant conventional datasets.
Our dataset is based on PromptTTS++ and provides human-annotated speaker characteristics for all speakers from LibriTTS-R\footnote{
We excluded some speakers with significant corrupted characteristics, such as male speakers altered to sound female. 
}, whereas PromptTTS++ offers annotations only for a limited portion of the entire LibriTTS-R dataset.

We provide two types of prompts: a \textit{style prompt} and a \textit{speaker prompt}.
The style prompt characterizes the speaking style of utterances and is defined for each utterance within the dataset. 
Specifically, it includes attributes such as gender, speaking speed, pitch, and loudness, following PromptSpeech~\cite{guo2023prompttts}. 
In contrast, the speaker prompt captures human perception of speaker characteristics consistent across different utterances and unique to each speaker in the dataset.

We adopt a hybrid approach to construct prompt annotations: (1) manual annotations of speaker prompts and (2) synthetic annotations of style prompts.
For the style prompts, we first analyze the statistics of fundamental frequency ($F_0$), number of syllables per second, and loudness for the entire dataset.
Then, we automatically label each attribute with five possible labels using the data statistics.
Finally, we utilize a set of pre-defined template sentences to map the set of discrete labels into natural language descriptions, followed by data augmentation by a large language model (LLM).
For the speaker prompts, we ask human annotators to collect perceptual and impression words for each speaker in the dataset.
Note that although our dataset is originally designed for prompt-based TTS, it can be used for novel tasks such as style captioning~\cite{yamauchi2024stylecap}.

\ryedit{We perform two experiments with our corpus: prompt-based TTS and style captioning~\cite{yamauchi2024stylecap}.}
Experimental results for prompt-based TTS show that the TTS model using LibriTTS-P achieves higher naturalness than the TTS model using PromptSpeech.
Furthermore, experimental results for style captioning tasks show that the model trained with LibriTTS-P generates 2.5 times more accurate words than the model with PromptSpeech.
Audio samples and style captioning examples are available on our demo page\footnote{\url{https://masayakawamura.github.io/libritts-p/}}.
The URL of our dataset will be included in the final version of our paper.

\begin{table}[t]
  \centering
  \caption{\ryedit{Comparison of LibriTTS-P with other datasets. Note that the number of prompts for PromptTTS++ indicates the count of samples for which both speaker and style prompts are available.}}
  \vspace{-2mm}
  \scalebox{0.90}{
  \begin{tabular}{c|c|c|c|c}
    \toprule
    Dataset & \# prompts & \# spks & \begin{tabular}[c]{@{}c@{}}Style\\ prompt\end{tabular} & \begin{tabular}[c]{@{}c@{}}Speaker\\ prompt\end{tabular} \\ \midrule
    PromptSpeech (Real) & 26,588 & 1,191 & \checkmark &  \\ 
    PromptTTS++ & 59,252 & 404 & \checkmark & \checkmark \\ 
    \textbf{LibriTTS-P (ours)} & 373,868 & 2,443 & \checkmark & \checkmark \\ 
    \bottomrule
  \end{tabular}
  }
  \vspace{-2mm}
  \label{tab:dataset_comparison}
\end{table}

\begin{table*}[ht!]
  \centering

  \caption{Example prompts: the underlined text indicates the style prompts, while the italicized text represents the speaker prompts.}
  \vspace{-2mm}  
  \scalebox{1.0}{
  \begin{tabularx}{\textwidth}{@{}>{\hsize=.25\hsize}X|>{\hsize=.3\hsize}X|>{\hsize=1.2\hsize}X@{}}
  \toprule
   Utterance ID & PromptSpeech~\cite{guo2023prompttts} & \textbf{LibriTTS-P (ours)} \\ \midrule
   1382\_130549\newline\_000092\_000000&Her sound height is really high, the volume is normal, but she speaks very slowly  & \underline{A woman speaks slowly, with a high-pitched voice and low volume.} \textit{The speaker's identity can be described as feminine, adult-like, slightly relaxed, soft, raspy, intellectual, calm, friendly, reassuring, slightly kind, modest.} \\ \hline
   1460\_138290\newline\_000011\_000004&Use her bass to say & \underline{Ask a woman to speak with normal pitch, normal speaking speed and standard volume.} \textit{Descriptions of the speaker's vocal style are feminine, adult-like, tensed, clear, fluent, intellectual, calm, slightly friendly, elegant, lively, slightly strict, slightly sharp.} \\
   \bottomrule
  \end{tabularx}
  }
  \vspace{-2mm}  
  \label{tab:exmaple_prompt}
\end{table*}

\section{LibriTTS-P}

Our corpus contains two types of prompts: a speaker prompt and a style prompt based on LibriTTS-R. Details are as follows.

\subsection{Speaker prompt}
To create speaker prompts, we manually annotate human perceptions of speaker characteristics for each speaker in the dataset.
Since performing automatic annotations on human perceptions is inherently challenging, we \kkkedit{ask} three professional annotators for this task.
To simplify the labeling process, we present two sets of special words associated with speaker identity to annotators: perception and impression words~\cite{kido1999voice, yamato2016toshiba}.
The perception words describe the perception of speech attributes such as gender and strength of the voice, whereas the impression words describe more subjective impressions of speech, such as cool and cute.
The list of the selected words can be found in Figure~\ref{fig:histogram_df}.
Note that even though we separate those words into two different categories, we treat them equally when used for training data.

In more detail, we first exclude audio files from LibriTTS-R where speech restoration failed.
This exclusion is due to the potential alteration of speaker identity caused by failed speech restoration during the creation of LibriTTS-R.
Then, we select audio files of less than 10 seconds in length for a given speaker from LibriTTS-R. 
From these selected files, the five longest are presented to the annotators.
All audio files are presented if less than five audio files are available.
The annotators listen to these audio files and select at least one applicable perceptual or impression word.
Additionally, each word is assigned a degree of intensity across three levels (e.g., slightly cute, cute, and very cute).
These selected words represent the speaker's identity, and this annotation process is applied to all speakers in LibriTTS-R.
We simply concatenate annotated words to create a speaker prompt.

\subsection{Style prompt}

As for style prompts, we perform synthetic annotations on pitch, speaking speed, and loudness.
First, for each utterance in the dataset, we calculate pitch, speaking speed, and loudness as the average $F_0$, the number of syllables per second, and loudness units relative to full scale~\cite{BS2011}, respectively.
Subsequently, we assign each utterance one of five levels (very-low, low, normal, high, and very-high) based on the statistics of those calculated metrics.
This classification \ryedit{is} performed using the quartile points of the distribution for each characteristic across all data.
Specifically, the bottom 10\% of the distribution is defined as very-low, the next 20\% as low, the middle 40\% as normal, the following 20\% as high, and the top 10\% as very-high for each characteristic.
To construct the style prompt with those assigned labels, we utilize a set of pre-defined template sentences to map the set of labels to natural language descriptions. 
We create the pre-defined template sentences by extending the style prompts in PromptSpeech~\cite{guo2023prompttts} with additional manual corrections.

\subsection{Data augmentation}
\label{ssec:dataaug}

We apply data augmentation to increase the diversity of the style prompts and speaker prompts.
For the style prompts, we employ Llama2~\cite{touvron2023llama2} for sentence rephrasing, creating 1,347 unique prompts.
For the speaker prompts, we simply use sentence templates.
Table~\ref{tab:exmaple_prompt} presents examples of prompts from PromptSpeech and LibriTTS-P, using templates such as \textit{"The speaker's identity can be described as"} and \textit{"Descriptions of the speaker's vocal style are"}.

\subsection{Dataset analysis}
\label{ssec:analysis}

\begin{figure}[t]
  \centering
  \includegraphics[width=\linewidth]{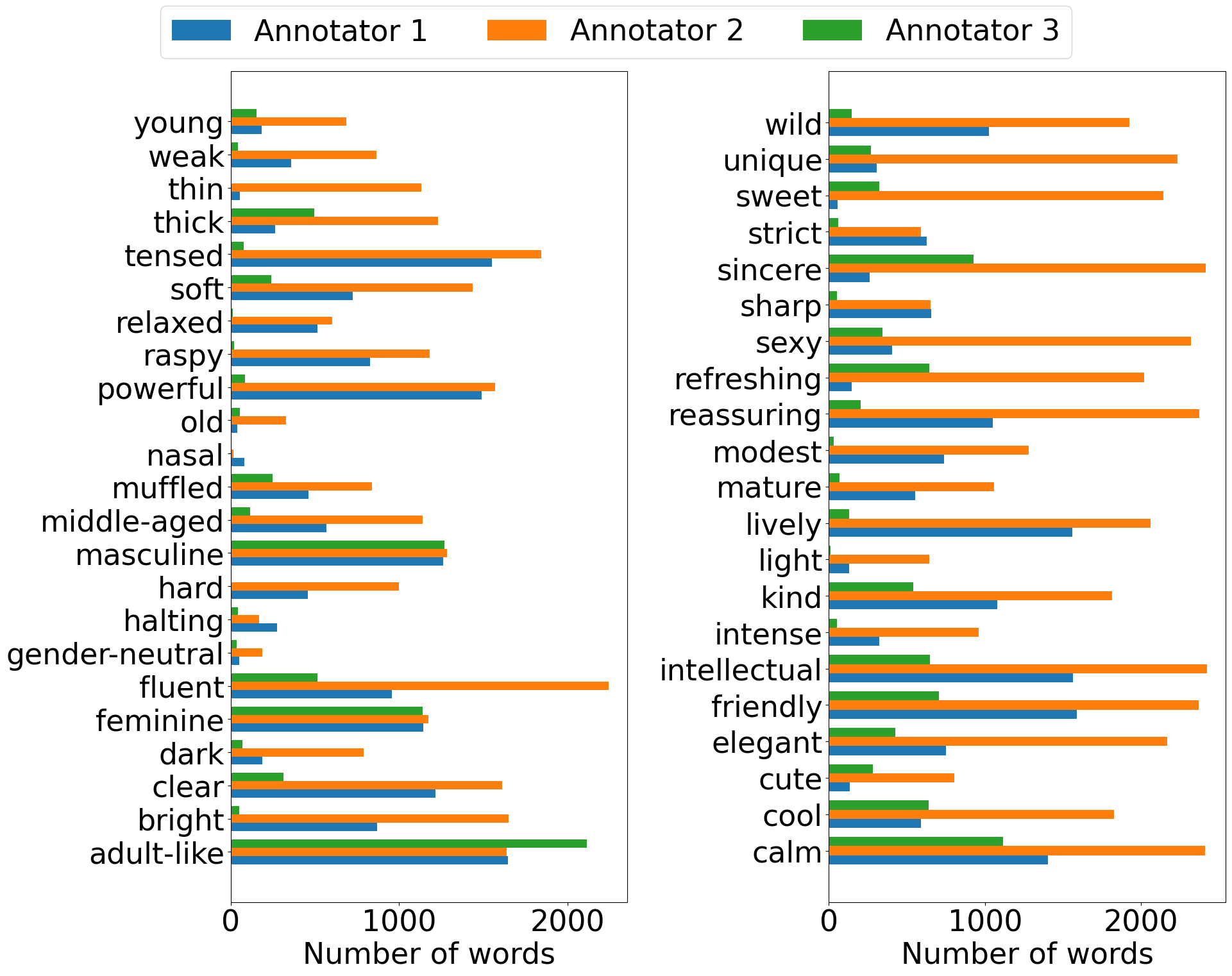}

  \caption{Histograms of \kedit{perceptual} \ryedit{words (left) and \kedit{impression} words (right)} for each annotator.}
  \label{fig:histogram_df}
  \vspace{-6mm}
\end{figure}

Figure~\ref{fig:histogram_df} shows the number of impression and perceptual words selected by each annotator.
While a nearly \hedit{equal} number of words were chosen for \textit{feminine} and \textit{masculine}, there was variability in the number of the other words representing speaker identity across the annotators.

To conduct a more detailed analysis, we evaluated the annotators' degree of agreement in \rrryedit{their use of} impression and perceptual words.
Table~\ref{tab:jaccard_similarity} shows the Jaccard \kedit{index}~\cite{jaccard1912} for each annotator \rrryedit{pair}, with higher values indicating that \rrryedit{two} annotators selected more similar perceptual and impression words. 
\rrryedit{First, we observed that} the Jaccard index \rrryedit{increased} for all the annotator pairs when adverbs of degree, \rrryedit{such as \textit{slightly} and \textit{very}, were ignored (e.g., treating \textit{very cute} as simply \textit{cute})}. 
This implies that the annotations of impression and perceptual words are more consistent when the intensity of each attribute is not considered. 
In addition, \rrryedit{we can see that the Jaccard index was lower when comparing data between annotator 2 and annotator 3 than with other pairs. This suggests that perceptions of speaker characteristics vary significantly among individuals; specifically, annotators 2 and 3 perceived the speakers more differently than other pairs, leading to a greater discrepancy in their use of perceptual and impression words.}
\rrryedit{Exploring this phenomenon through larger-scale human annotations could be valuable; however,} we leave it to our future work.

\begin{table}[t]
  \centering
  \caption{\kkkedit{Jaccard \kedit{index} of speaker prompts among annotators. \textbf{w/ adverb} and \textbf{w/o adverb} in the table \kkkedit{denote} adverbs of degree, such as slightly and very, which are considered and ignored, respectively.}}
  \vspace{-2mm}
  \scalebox{1.0}{  
  \begin{tabular}{c|cc}
    \toprule
    Annotators & w/ adverb &  w/o adverb \\ \midrule
    Annot. 1 / Annot. 2 & 0.125 & 0.370 \\
    Annot. 2 / Annot. 3 & 0.057 & 0.215 \\
    Annot. 3 / Annot. 1 & 0.151 & 0.258 \\ 
    \bottomrule
  \end{tabular}
  }
  \vspace{-0mm}
  \label{tab:jaccard_similarity}
\end{table}

\begin{table}[t]
  \centering
  \caption{Naturalness and audio-prompt consistency MOS \kedit{with 95\% confidence intervals}. Averaged scores 
  among 10 speakers in the evaluation data were reported.}
  \vspace{-2mm}
  \begin{tabular}{@{}c|cc@{}}
  \toprule
   \multirow{2}{*}{Dataset} & \multirow{2}{*}{Naturalness MOS} & Audio-prompt \\
   & & consistency MOS \\ \midrule
    PromptSpeech & $2.47\pm0.12$ & {$3.06\pm0.10$} \\
    LibriTTS-P & \bm{$3.37\pm0.13$} & \bm{$3.11\pm0.10$} \\ \midrule
    Ground truth & $4.58\pm0.09$ & $3.22\pm0.09$ \\ 
    \bottomrule
  \end{tabular}
  \label{tab:result_promptttspp}
\end{table}

\begin{table*}[ht!]
  \centering
  \caption{Results of StyleCap trained with PromptSpeech and LibriTTS-P. The terms \# accurate words and \# inaccurate words represent the average number of words that accurately and inaccurately describe the speech \kedit{with 95\% confidence intervals}, respectively. \# words are the average number of words per generated prompt.
  }
  \vspace{-2mm}
  \begin{tabularx}{\textwidth}{@{}l|XXp{2cm}p{2cm}p{2cm}@{}}
  \toprule
    & \multicolumn{2}{c}{Subjective evaluation} & \multicolumn{3}{c}{Objective evaluation}\\ \cmidrule(lr){2-3} \cmidrule(lr){4-6}
   Dataset & \# accurate words$\uparrow$ & \# inaccurate words$\downarrow$ & BLEU@4$\uparrow$ & BERT-Score$\uparrow$ & \# words\\ \midrule
    PromptSpeech  & $2.09\pm{0.10}$ & \bm{$0.66\pm{0.080}$} & $0.17$ & $0.88$ & $11.42$\\ 
    LibriTTS-P & \bm{$5.24\pm{0.20}$} & \bm{$0.66\pm{0.080}$}& \bm{$0.68$} & \bm{$0.95$} & \bm{$25.56$} \\ 
   \bottomrule
  \end{tabularx}
  \vspace{-4mm}
  \label{tab:result_stylecap}
\end{table*}

\section{Experimental evaluations}

We utilized the LibriTTS-P dataset for prompt-based controllable TTS and style captioning as example applications and subsequently assessed the dataset's effectiveness.


\subsection{Prompt-based controllable TTS}
\subsubsection{Experimental conditions}

We adopt PromptTTS++~\cite{shimizu2024prompttts} for the TTS experiments, as it has demonstrated its capability of prompt-based speaker identity control.
We separately trained two PromptTTS++ models with PromptSpeech~\cite{guo2023prompttts} and LibriTTS-P for comparison.
When trained with PromptSpeech, PromptTTS++ utilized its style prompts. In training with LibriTTS-P, PromptTTS++ used both its speaker and style prompts. For both models, the corresponding speech data from LibriTTS-R was used~\cite{koizumi2023librittsr}.
For evaluation data, we \kedit{selected} 10 speakers, specifically with the speaker IDs: 121, 237, 260, 908, 1089, 1188, 1284, 1580, 1995, and 2300.
The rest of each dataset was split into training and validation data, with the split based on 2\% for validation.
For training the duration predictor in PromptTTS++, we extracted phoneme duration using the Montreal forced aligner~\cite{mcauliffe2017montreal}.
Most of the hyperparameters followed the original PromptTTS++ paper~\cite{shimizu2024prompttts}.
We trained the PromptTTS++ models using each dataset for 100~epochs and used AdamW optimizer~\cite{los2019adamw} with a batch size of 30~K~frames.
We adopted a warmup learning rate scheduler~\cite{vaswani2017attention} with an initial learning rate of 0.001. 
The number of warmup steps was set to 4000.
We used BigVGAN-base~\cite{lee2023bigvgan} vocoder trained on LibriTTS-R~\cite{koizumi2023librittsr}. 
To stabilize the pitch generation, \kedit{the} source excitation module from the neural source filter model~\cite{wang2020neural} was introduced in this vocoder.
The BigVGAN vocoder was trained for 2.5~M~steps with a batch size of 32 using the AdamW optimizer.
\subsubsection{Evaluations}

We performed two subjective evaluations: 5-point naturalness mean opinion score (MOS) and 4-point \kedit{audio-prompt consistency} MOS tests.
For the naturalness MOS test, human raters are asked to check the quality of the audio samples using the following five possible responses: 1 = Bad; 2 = Poor; 3 = Fair; 4 = Good; and 5 =
Excellent.
For the audio-prompt consistency MOS test, raters are asked to \kedit{judge} the consistency between the audio samples and the corresponding prompts using the following four possible responses: 1 = Inconsistent; 2 = Somewhat inconsistent; 3 = Somewhat consistent; and 4 = Consistent.
For each test speaker from LibriTTS-R, we randomly selected three utterances. We then synthesized speech samples using the corresponding prompts and transcriptions as input. 
Note that we used 15 style prompts from PromptSpeech and 15 speaker/style prompts from LibriTTS-P.
In total, we evaluated 30 utterances for each model.
We asked 10 participants for the MOS tests.

Table~\ref{tab:result_promptttspp} shows the subjective evaluation results.
As for naturalness, the model trained with LibriTTS-P achieved a significantly higher score than the model trained with PromptSpeech.
Both models exhibited comparable performance in terms of audio-prompt consistency, though the model trained with LibriTTS-P achieved a marginally higher score.
Note that the audio-prompt consistency MOS showed no significant difference between LibriTTS-P and ground truth according to the results of a student's $t$-test at a 5\% significance level. 
These results demonstrated that our dataset enables more high-quality prompt-based controllable TTS systems.


\subsection{Style captioning}

\subsubsection{Experimental conditions}

We adopt StyleCap~\cite{yamauchi2024stylecap} for style captioning experiments.
We trained two StyleCap models separately with PromptSpeech~\cite{guo2023prompttts} and LibriTTS-P.
The model comprises a speech encoder, a mapping network, and a text decoder.
The speech encoder is composed of a feature extractor backbone and aggregation module.
We used pre-trained WavLM BASE+~\footnote{\url{https://huggingface.co/microsoft/wavlm-base-plus}}~\cite{chen2022wavlm} as the feature extractor backbone.
The aggregation module consists of a stack of bidirectional long short-term memory~\cite{hochreiter1997long} and multi-head attention~\cite{vaswani2017attention}.
As the text decoder, we used a pre-trained GPT-2 model~\footnote{\url{https://huggingface.co/openai-community/gpt2}}~\cite{radford2019language}.
We used 90\% of each dataset for training, with the remaining data equally divided between validation and evaluation data.
We trained the StyleCap \ryedit{models} using each dataset for 20 epochs with AdamW optimizer~\cite{los2019adamw}. 
We set the average batch sizes to 11 and 10 when training with PromptSpeech and LibriTTS-P, respectively.
We adopted a warmup learning rate scheduler~\cite{vaswani2017attention} with an initial learning rate of 0.00002. 
The number of warmup steps was set to 5000.
The other experimental conditions are the same as the original paper~\cite{yamauchi2024stylecap}.

We conducted both subjective and objective evaluations to evaluate the performance of the StyleCap models.
For subjective evaluation, we measured \ryedit{audio-prompt consistency} between the generated prompts and the speech. 
In detail, we first used \kkkedit{GPT-4}~\cite{achiam2023gpt4} to extract words from the prompts that are relevant to speaking style and speaker identity, using the instruction: \textit{“Please extract as many terms as possible that represent speaker identity and speech style”}.
For instance, given the prompt \textit{“A woman is asked to speak slowly with low pitch and normal volume. Descriptions of the speaker's vocal style are feminine, adult-like.”}, 
we can \rrryedit{extract} the \rrryedit{following} terms: \textit{“woman, slowly, low pitch, normal volume, feminine, adult-like”}, which facilitates the evaluation process by providing a quantifiable number of accurate words.
Then, we asked 10 human raters to assess 30 randomly sampled audio samples and their corresponding prompts, posing the two questions:
(Q1) \textit{Which words do you feel are suitable to represent the speech?} (Q2) \textit{Which words do you feel are unsuitable to represent the speech?}
These two questions are meant to evaluate the accuracy and variety of the generated prompts.

For objective evaluations, we used bilingual evaluation understudy (BLEU) @4~\cite{papineni2002bleu} and BERT-Score~\footnote{\url{https://github.com/Tiiiger/bert_score}}~\cite{zhang2020bertscore}.
BLEU@4 focuses on the overlap between the generated prompts and the ground truth prompts up to $4$-gram.
To calculate BLEU, we utilized the natural language toolkit~\footnote{\url{https://github.com/nltk/nltk}}~\cite{bird2009natural}.
BERT-Score is based on pre-trained BERT embeddings to represent and match the tokens in the ground truth and generated prompts.
By computing the similarity between the generated prompts and the ground truth prompts, BERT-Score can better capture \rrryedit{their semantic similarity}.
We also computed the number of words in the generated prompts.
To calculate these scores, we randomly selected 250 samples from the evaluation data.
For the model trained with PromptSpeech, we used the corresponding evaluation dataset from PromptSpeech as the reference.
Similarly, for the model trained with LibriTTS-P, we used the LibriTTS-P evaluation dataset as the reference.
Note that the reference prompts differed between PromptSpeech and LibriTTS-P, meaning that the BLEU and BERT-Score results may not be directly comparable.

\subsubsection{Evaluations}
Table~\ref{tab:result_stylecap} shows the results of \rrryedit{the} subjective and objective evaluations.
In the subjective evaluation, the model trained with LibriTTS-P generated prompts with a number of accurate words that was 2.5 times greater than that of the model trained with PromptSpeech. Furthermore, the number of inaccurate words in the prompts from the model trained with LibriTTS-P was comparable to that of the model trained with PromptSpeech. 
The results suggest that employing LibriTTS-P, which includes both speaker and style prompts, facilitates the generation of more diverse prompts without compromising accuracy.
The objective evaluation results also confirmed the superior performance of the model trained with LibriTTS-P in terms of BLEU@4, BERT-Score, and the number of words.
We hypothesize the two primary factors contributed to the superior performance: (1) the larger size of the LibriTTS-P dataset compared to PromptSpeech, and (2) the incorporation of speaker prompts in LibriTTS-P, which provides more detailed information about speech.

\section{Discussions}

Our experiments on prompt-based TTS and style captioning have demonstrated the capabilities of our new corpus. 
Nevertheless, several challenges persist.
As analyzed in Section~\ref{ssec:analysis}, an expansion of large-scale human annotations is necessary to cover the diverse characteristics of human perception.
Moreover, investigating the use of free-form text descriptions for their flexibility represents an important research direction, as seen in NLPSpeech~\cite{yang2023instructtts} and Coco-Nut~\cite{watanabe2023coco-nut}.
For example, while our hybrid annotation approach efficiently scales by combining automatic and manual methods, it cannot capture the dynamic nature of speech, e.g., the following prompt cannot be obtained: ``\textit{a woman whispers slowly at the beginning but turns to speak fast in a relaxed tone at the end}".
Employing free-form descriptions could mitigate this limitation, albeit at a higher annotation cost.
Finally, despite the substantial size of LibriTTS-R (i.e., 585 hours), we observed that its range of styles and speaker diversity might be inadequate; for instance, there is a lack of variation in energy levels. 
Exploring larger datasets could also be a valuable avenue for future research.

\section{Conclusion}
In this paper, we introduced LibriTTS-P, a new dataset that includes style and speaker prompts based on a high-quality English corpus LibriTTS-R.
Unlike conventional public English prompt datasets such as PromptSpeech, our dataset incorporates speaker prompts that capture human perceptions of speaker characteristics.
We demonstrated the capabilities of our corpus with prompt-based TTS and style captioning.
We believe that LibriTTS-P accelerates the development of prompt-based TTS, style captioning, and other \rrryedit{new} applications.

\bibliographystyle{IEEEtran}
\bibliography{mybib}

\end{document}